\documentclass[aip,
nofootinbib,
rsi,%
reprint,
twocolumn,
longbibliography,
author-numerical%
]{revtex4-2}

\usepackage[utf8]{inputenc}
\usepackage[T2A]{fontenc}
\usepackage{graphicx}
\usepackage{textcomp}

\usepackage{amsfonts}
\usepackage{amssymb}
\usepackage{amsmath}
\usepackage{color}

\usepackage{changes}

\newcommand{\pdr}[2]{\dfrac{\partial {#1}}{\partial #2}}

\newcommand{\pddr}[2]{\dfrac{\partial^2{#1}}{\partial #2^2}}

\newcommand{\pddra}[2]{{\partial^2 #1}/{\partial #2^2}}

\newcommand{\tx}{\tilde{x}}

\newcommand{\tz}{\tilde{z}}

\newcommand{\tit}{\tilde{t}}
\newcommand{\td}{\tilde{d}}

\newcommand{\tom}{\widetilde{\omega}}

\newcommand{\tj}{\tilde{j}}
\newcommand{\tk}{\tilde{k}}
\newcommand{\tc}{\tilde{c}}
\newcommand{\tp}{\tilde{p}}
\newcommand{\teta}{\tilde{\eta}}

\newcommand{\tJ}{\widetilde{J}}

\newcommand{\tZ}{\widetilde{Z}}
\newcommand{\tD}{\widetilde{D}}
\newcommand{\tN}{\widetilde{N}}
\newcommand{\tl}{\tilde{l}}

\newcommand{\tR}{\widetilde{R}}

\newcommand{\expo}{{{\rm e}^{\teta_0}}}

\newcommand{\cref}{c_{ref}}

\newcommand{\jlim}{j_{\lim}}

\newcommand{\Cdlast}{C_{dl}^*}
\newcommand{\Cdl}{C_{dl}}
\newcommand{\ksig}{\tk_{\sigma}}
\newcommand{\ks}{\tk_s}
\newcommand{\kD}{\tk_D}

\newcommand{\tjlim}{\tj_{\lim}}

\newcommand{\tCdl}{\widetilde{C}_{dl}}

\newcommand{\lam}{\lambda}

\newcommand{\veps}{\varepsilon}

\newcommand{\sion}{\sigma_p}
\newcommand{\tsion}{\tilde{\sigma}_p}
\newcommand{\sionast}{\sigma_p^*}

\newcommand{\Dox}{D_{ox}}
\newcommand{\tDox}{\tD_{ox}}
\newcommand{\tDoxd}{\tD_{ox,d}^0}

\newcommand{\Doxdry}{D_{ox,dry}}

\newcommand{\matr}[1]{\mathbf{#1}}
\newcommand{\matU}{\matr{U}}
\newcommand{\matM}{\matr{M}}
\newcommand{\matA}{\matr{A}}

\newcommand{\vecv}{\matr{v}}

\newcommand{\lcat}{l_t}

\newcommand{\ri}{{\rm i}}

\newcommand{\eeo}{{\rm e}^{\teta^0}}

\newcommand{\lexp}[1]{\exp\left(#1\right)}

\begin{document}


\sf

\title{A fast two-phase model for PEM fuel cell impedance}

\author{Andrei Kulikovsky}
\email{A.Kulikovsky@fz-juelich.de}

\affiliation{Forschungszentrum J\"ulich GmbH             \\
    Theory and Computation of Energy Materials (IET-3)   \\
    Institute of Energy and Climate Research,            \\
    D-52425 J\"ulich, Germany
}

\begin{abstract}
We report a physics-based matrix model for PEM fuel cell impedance that
takes into account liquid water transients in the cathode catalyst layer (CCL).
This model is an extension of our previous single-phase matrix model for functionally graded CCLs.
The solution to the boundary-value problem for the perturbation amplitudes
is constructed from basic solutions to the respective Cauchy problem with constant
coefficients on numerical cells. The single-phase model employed analytical solutions
to this problem. Here, for a larger system of ODEs, the analytical solution does not exist and
we used spectral decomposition of
the ODE system matrix to obtain the basic solutions numerically.
Our impedance model is two orders of magnitude faster than a model based
on a standard, general-purpose boundary-value solver. Fitting the matrix model to an experimental spectrum
takes about a minute on a laptop. The developed matrix method
can be used to numerically solve high-order linear boundary-value problems with variable coefficients.
\end{abstract}


\maketitle

\newpage

\section{Introduction}

Electrochemical impedance spectroscopy (EIS) allows us to characterize PEM
fuel cells in situ without interrupting power production. This makes EIS
unique among other characterization techniques.
Applying a small probing AC signal to the operating cell
does not change its status, but provides a lot of information on the transport
and kinetic processes inside the cell.

Measuring impedance spectra is a routine procedure that can even be carried out
on board a PEMFC-based vehicle. However, interpreting the spectra is a non-trivial task.
A simple and widely used approach is to construct an equivalent electric circuit
with the elements representing basic processes inside the cell \cite{Han_10,Kim_13,Minggao_21}.
These elements can
include resistances, capacitances, constant phase elements, inductive coils,
as well as analytical solution of specific problems,
such as Warburg finite-length impedance. Fitting the constructed circuit to
the experimental spectrum is fast, and can estimate the contribution of
each element to cell voltage loss. However, none of the equivalent
circuits is unique, which may raise concerns about its relevance \cite{Macdonald_06}.

In recent years, the distribution of relaxation times (DRT) method became popular.
This method is based on the formal expansion of the impedance spectrum over an infinite
series of parallel RC circuits \cite{Fuoss_41,Schlichlein_02,Tsur_11,Wan_15,Boukamp_15}.
Each process inside the cell has its own ``resonance'' frequency.
The DRT expansion is expected to reveal a number of peaks on the frequency scale,
with each peak representing a single process within the cell.
However, for oxygen transport processes, the RC kernel used for the DRT expansion
is not relevant, leading to phantom peaks. Furthermore,
the characteristic frequencies of two or more processes may overlap,
making them indistinguishable in the DRT spectrum. Lastly, low-frequency ``inductive'' loops
in Nyquist spectra cannot be expanded into an RC series.
Nevertheless, the DRT spectrum provides a useful ``snapshot'' of the cell state.

The most reliable method for the spectra analysis
is the physics-based modeling (see reviews \cite{Zhang_20,Huang_20}).
Experimental PEMFC spectra often exhibit a low-frequency ``inductive'' loop \cite{Pivac_16}.
One reason for the formation of such a loop is the effect of liquid water transients
on the cathode catalyst layer (CCL) proton conductivity \cite{Kulikovsky_22h,Kulikovsky_24b}.
So far, very few studies have incorporated two-phase effects into
PEMFC impedance models \cite{Bao_15,Futter_18,Kulikovsky_24b,Kulikovsky_24e,Weber_26}.
In \cite{Bao_15,Futter_18,Weber_26}, a qualitative resemblance of the model and experimental spectra
has been demonstrated. Two-phase models suitable for the least-squares
fitting of the experimental spectra have been reported in\cite{Kulikovsky_24b,Kulikovsky_24e}.

Generally, any transient model for the cell performance
can be converted into the impedance model by means of the standard procedure of
linearization and Fourier-transform. In this way, a one-dimensional transient model
for the through-plane transport processes in the cell
reduces to a boundary-value problem (BVP) for the system of linear complex-valued ODEs.
This system can be linked in a 1d+1d manner to the problem of oxygen transport
in the cathode channel. This approach enables incorporation of the most important
processes contributing to the cell impedance \cite{Kulikovsky_16b,Kulikovsky_24e}.

However, even a single-phase model leads to a BVP that
includes two second-order linear ODEs for proton and oxygen transport in the CCL.
The least-squares fitting codes require this BVP to be
solved hundreds of times while the algorithm searches for the optimal fitting parameters.
This procedure is time-consuming, taking up to an hour on a standard laptop.
With the growth of the model complexity, this time increases dramatically.

Below, we develop a fast, physics-based matrix model for cell impedance
that incorporates liquid water transport and evaporation in the CCL.
This model is suitable for cells with a functionally graded catalyst layer.
Our previous single-phase matrix model was based on  analytical solutions
for the oxygen and proton transport perturbations in an elementary computational cell
\cite{Kulikovsky_24f,Kulikovsky_26h_arxiv}.
These solutions were used to construct the transform matrix relating the state vectors on
the either side of the CCL. Here, however, in each computational cell, we are dealing with a system
of three second-order, complex-valued ODEs with constant coefficients,
which has no analytical solution. To solve this system and to construct
the transfer matrix, we use the spectral decomposition of the ODEs coefficient matrix.

The resulting impedance model is two orders of magnitude faster than the reference model based
on the standard, general-purpose BVP solver. Fitting the matrix model to an experimental
impedance spectrum takes about a minute on a laptop. This makes the fitting of the matrix
model to experimental spectra competitive with the equivalent circuit approach, while retaining
all the advantages of a physics-based description of the cell impedance.

\section{Model}

\subsection{Basic equations}

The core of the two-phase impedance model is the system of conservation
equations for the proton current, oxygen concentration, liquid saturation,
and water vapor  in the CCL \cite{Eikerling_06a,Kulikovsky_24b}:
\begin{align}
    &C_{dl}\pdr{\eta}{t} - \pdr{}{x}\left(\sion \pdr{\eta}{x}\right) = - R_{ORR}
	\label{eq:etax} \\
    & \pdr{c}{t} - \pdr{}{x}\left(\Dox\pdr{c}{x}\right) = - \dfrac{R_{ORR}}{4F},
	\label{eq:cx}   \\
    &\dfrac{1}{V_{m}}\pdr{s}{t} - \dfrac{K_L}{\mu_w V_m}\pddr{p_L}{x}
        = \dfrac{2R_{ORR}}{4F} - \dfrac{2R_{ev}}{4F},
	\label{eq:pLx} \\
    & \pdr{c_v}{t} - D_v \pddr{c_v}{x} = \dfrac{2R_{ev}}{4F},
    \label{eq:pvx}
\end{align}
where $R_{ORR}$ is the Tafel ORR rate (A~cm$^{-3}$)
\begin{equation}
    R_{ORR} = i_*\left(\dfrac{c}{\cref} \right)\lexp{\dfrac{\eta}{b_T}}.
    \label{eq:Rorr}
\end{equation}
and $R_{ev}$ is the rate of water evaporation
\begin{equation}
   R_{ev} = \dfrac{2e_0 \kappa \xi_{ev} R T}{\lcat} (c_v^s - c_v)
   \label{eq:Rec}
\end{equation}
Here,
$t$ is time,
$x$ is the distance from the membrane,
$\Cdl$ is the double layer capacitance,
$\eta$ is the positive by convention ORR overpotential,
$\sion$ is the CCL proton conductivity,
$b_T$  is the ORR Tafel slope,
$c$ is the oxygen concentration,
$\cref$ is the reference (inlet) oxygen concentration,
$\Dox$ is the CCL oxygen diffusivity,
$s$ is the liquid saturation,
$v_m$ is the water molar volume,
$K_L$ is the absolute liquid hydraulic permeability of the CCL,
$\mu_w$ is the water dynamic viscosity,
$p_L$ in the liquid pressure,
$c_v$ is the water vapor concentration,
$i_*$ is the ORR exchange current density,
$c_v^s$ is the saturation water vapor concentration,
$e_0$ is the elementary charge,
$\kappa$ is the evaporation rate constant,
$\xi_{ev}$ is the evaporation interfacial area factor.

The CCL proton conductivity $\sion$ and oxygen diffusivity $\Dox$ depend
on the liquid saturation $s$ according to\cite{Fathi_17,Springer_91}
\begin{equation}
   \Dox = (1 - s)^2 \Doxdry, \quad \sion = k_{\sigma} s
   \label{eq:Doxs}
\end{equation}
where $\Doxdry$ is the oxygen diffusivity of the ``dry'' CCL and
$k_{\sigma}$ is constant.
The liquid saturation is related to the capillary pressure by the
water retention curve\cite{Olbrich_22}:
\begin{equation}
   s = s_* + \dfrac{(1 - s_*)}{2}\left(1 + \tanh\left(\dfrac{p_c - p_{c,0}}{k_{pc}} \right) \right)
   \label{eq:wrt}
\end{equation}
Here,
$s_*$ is the minimal static liquid saturation,
$p_c = p_L - p_g$ is the capillary pressure,
$p_g$ is the gaseous pressure,
$p_{c,0}$ and $k_{pc}$ describe
the position and slope of the transition region, respectively.
Eq.\eqref{eq:wrt} describes a
smooth transition from $s= s_*$ to the fully saturated state with $s=1$.

Oxygen transport in the GDL obeys the Fick's diffusion equation
\begin{equation}
   \pdr{c_b}{t} - D_b \pddr{c_b}{x} = 0
   \label{eq:cbx}
\end{equation}
where
$c_b$ is the oxygen concentration in the GDL,
$D_b$ is the GDL oxygen diffusivity.

Lastly, the oxygen transport in the channel is described by the plug-flow equation
\begin{equation}
   \pdr{c_h}{t} + v\pdr{c_h}{z} = - \left.\dfrac{D_b}{h}\pdr{c_b}{x}\right|_{x=l_t + l_b}
   \label{eq:chz}
\end{equation}
where
$c_h$ is the oxygen concentration in the channel,
$v$ is the constant flow velocity,
$z$ is the coordinate along the channel,
$h$ is the channel height,
$l_b$ is the GDL thickness,
and the right side describes the oxygen sink in the perpendicular direction
through the GDL.

Detailed description of the model \eqref{eq:etax}--\eqref{eq:wrt} can be
found in\cite{Kulikovsky_24b}. The coupled 1d+1d solutions to Eqs.\eqref{eq:cbx} and
\eqref{eq:chz} are discussed below.

\subsection{Equations for perturbation amplitudes}

Below, the dimensionless variables will be used:
\begin{multline}
    \tit = \dfrac{t}{t_*} , \quad \tx = \dfrac{x}{\lcat},
    \quad \tc = \dfrac{c}{\cref}, \quad \teta = \dfrac{\eta}{b_T},
    \quad \tj = \dfrac{j}{j_*},  \\
    \tp = \dfrac{p}{\cref RT},  \quad
    \tCdl = \dfrac{\Cdl}{\Cdlast}, \quad \tsion = \dfrac{\sion}{\sionast},  \quad \tom = \omega t_*, \\
     \tD = \dfrac{4 F D \cref}{\sionast b_T}, \quad \tz = \dfrac{z}{L}, \quad
     \tZ = \dfrac{Z\sionast}{\lcat}.
    \label{eq:dless}
\end{multline}
where
\begin{equation}
   j_* = \dfrac{\sionast b_T}{\lcat}, \quad t_* = \dfrac{\Cdlast b_T}{i_*}
   \label{eq:ast}
\end{equation}
are the characteristic current density and time, respectively,
$\Cdlast$ and $\sionast$ are the characteristic DL capacitance and proton
conductivity,
$\lcat$  is the CCL thickness,
$\omega$ is the angular frequency of the applied AC perturbation,
$Z$ is the impedance, and  $L$ is the channel length.

\begin{figure}
\begin{center}
    \includegraphics[scale=0.8]{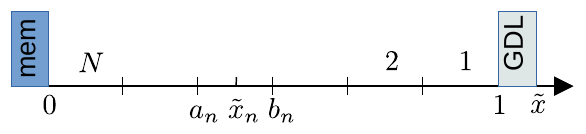}
\end{center}
    \caption{Schematic of the cathode catalyst layer separated into $N$
        numerical cells. The problem \eqref{eq:teta1x}--\eqref{eq:tpL1x}
        is solved in every cell $[a_n, b_n]$.
        Note that the cells are numbered from the GDL/CCL interface
        toward the membrane.
    }
    \label{fig:sketch}
\end{figure}

Linearization and Fourier-transform of equations \eqref{eq:etax}--\eqref{eq:pLx}
lead to the system of equations for the small perturbation amplitudes
$\teta^1$, $\tc^1$, and $\tp_L^1$, which can be written in a compact form  as
\begin{multline}
   \pdr{}{\tx}\left(\tsion^0 \pdr{\teta^1}{\tx}\right)
       = p_1 \teta^1 + q_1 \tc^1 + r_1 \tp_L^1
                     + u_1\pdr{\tc^1}{\tx} + w_1\pdr{\tp_L^1}{\tx}, \\
          \teta^1(1) = \teta_1^1, \quad \left.\pdr{\teta^1}{\tx}\right|_{\tx=1} = 0
       \label{eq:teta1x2}
\end{multline}
\begin{multline}
    \pdr{}{\tx}\left(\tDox^0  \pdr{\tc^1}{\tx}\right)
       = p_2 \teta^1 + q_2 \tc^1 + r_2 \tp_L^1 + u_2\pdr{\tc^1}{\tx} + w_2\pdr{\tp_L^1}{\tx}, \\
    \left.\pdr{\tc^1}{\tx}\right|_{\tx=0} = 0, \quad \tc^1(1) = \tc_b^1(1)
        \label{eq:tc1x2}
\end{multline}
\begin{multline}
    \tk_L\pddr{\tp_L^1}{\tx} = p_3 \teta^1 + q_3 \tc^1 + r_3 \tp_L^1, \\
    - \left.\tk_L\pdr{\tp_L^1}{\tx}\right|_{\tx=0} = \tN_{L,mem}^1, \quad \tp_L(1) = \tp_{cell}
        \label{eq:tpL1x2}
\end{multline}
Eqs.\eqref{eq:teta1x2}--\eqref{eq:tpL1x2}  form a linear
boundary-value problem.
Here,
$\teta_1^1$ is the amplitude of the applied potential perturbation,
$\tc_b^1$ is the oxygen concentration perturbation on the GDL side
of the CCL/GDL interface,
$\tN_{L,mem}^1$ is the perturbation of water flux to/from the membrane.
The coefficients in this system are given in Appendix \ref{sec:app}.
The superscripts 0 and 1 mark the steady-state variables and small
perturbation amplitudes, respectively.

The perturbation of the water vapor
concentration is zero, i.e. $\tc_v^1 = 0$, due to zero boundary conditions
at the membrane and CCL/GDL interfaces. For this reason, the equation for $\tc_v^1$
is omitted. However, the static value $\tc_v^0$ appears in the steady-state
equations.

The coefficients in Eqs.\eqref{eq:teta1x2}--\eqref{eq:tpL1x2} contain
the steady-state overpotential $\teta^0$,
oxygen concentration, $\tc^0$, water vapor concentration, $\tc_v^0$,
and liquid pressure, $\tp_L^0$ (Appendix~\ref{sec:app}).
These variables obey the static equations discussed in Appendix \ref{sec:ssp}.

\subsection{Matrix solution of the problem Eqs.\eqref{eq:teta1x2}--\eqref{eq:tpL1x2}}

We introduce a numerical mesh on the interval
$\tx \in [0,1]$ with $N$ cells numbered from the CCL/GDL interface ($\tx=1$) to the
membrane/CCL interface ($\tx=0$, Figure~\ref{fig:sketch}).
Following the idea of a matrix method, we consider the system
of equations \eqref{eq:teta1x2}--\eqref{eq:tpL1x2} on a cell
$\tx \in [a_n,b_n]$, assuming that the static transport parameters $\tsion^0$, $\tDox^0$
and all the coefficients in the system \eqref{eq:teta1x}--\eqref{eq:tpL1x}
are independent of $\tx$ on this cell.
However, the transport parameters and coefficients may vary between the numerical cells.
On $[a_n,b_n]$, the system \eqref{eq:teta1x2}--\eqref{eq:tpL1x2}, therefore, transforms to
\begin{equation}
   \tsion^0 \pddr{\teta^1}{\tx}
       = p_1 \teta^1 + q_1 \tc^1 + r_1 \tp_L^1 + u_1\pdr{\tc^1}{\tx} + w_1\pdr{\tp_L^1}{\tx} ,
       \label{eq:teta1x}
\end{equation}
\begin{equation}
    \tDox^0  \pddr{\tc^1}{\tx}
       = p_2 \teta^1 + q_2 \tc^1 + r_2 \tp_L^1 + u_2\pdr{\tc^1}{\tx} + w_2\pdr{\tp_L^1}{\tx},
        \label{eq:tc1x}
\end{equation}
\begin{equation}
    \tk_L  \pddr{\tp_L^1}{\tx} = p_3 \teta^1 + q_3 \tc^1 + r_3 \tp_L^1.
        \label{eq:tpL1x}
\end{equation}

Below, the subscript $n$ is omitted.
Further, on $[a, b]$, we consider the Cauchy problem
for Eqs.\eqref{eq:teta1x}--\eqref{eq:tpL1x} with the boundary conditions at $\tx=b$:
\begin{equation}
\begin{split}
   &\teta^1(b) = \teta_b^1,   \quad \left.   - \tsion^0\pdr{\teta^1}{\tx}\right|_{\tx=b} = \tN_{p,b}^1 \\
   &\tc^1(b)   = \tc_b^1,     \quad \left.   - \tDox^0 \pdr{\tc^1  }{\tx}\right|_{\tx=b} = \tN_{ox,b}^1 \\
   &\tp_L^1(b) = \tp_{L,b}^1, \quad \left.   - \tk_L   \pdr{\tp_L^1}{\tx}\right|_{\tx=b} = \tN_{L,b}^1
\end{split}
   \label{eq:bcxb}
\end{equation}
where $\tN_{p,b}^1$, $\tN_{ox,b}^1$, and  $\tN_{L,b}^1$ are the perturbations of the
proton, oxygen, and liquid water fluxes at $\tx=b$.

Let the state vector with the unknown functions and fluxes be
\begin{equation}
    \vecv = \left(\teta^1,\tN_p^1, \tc^1, \tN_{ox}^1, \tp_L^1, \tN_L^1 \right)^T
    \label{eq:vecv}
\end{equation}
Our nearest goal is to find the transform matrix $\matM_{ab}$ relating
the state vectors $\vecv_a$ at $\tx=a$ and $\vecv_b$ at $\tx=b$:
$\vecv_a = \matM_{ab}\vecv_b$.

Analytical solution to the problem \eqref{eq:teta1x}--\eqref{eq:tpL1x}
with the boundary conditions \eqref{eq:bcxb} does not exist.
However, there is an efficient
method for finding the transform matrix  numerically. Reformulating Eqs.\eqref{eq:teta1x}--\eqref{eq:tpL1x}
as a system of six first-order equations, we can write the problem in a matrix notation as
\begin{equation}
    \pdr{\vecv}{\tx} = \matA \vecv, \quad \vecv|_{\tx=b} = \vecv_b
    \label{eq:eqnA}
\end{equation}
where the matrix $\matA$ follows from Eqs.\eqref{eq:teta1x}--\eqref{eq:tpL1x}:
\begin{equation}
\matA =
\begin{pmatrix}
     0                   & -\dfrac{1}{\tsion^0}  & 0    & 0              & 0    & 0           \\
   - p_1                 & 0    & -q_1 & \dfrac{u_1}{\tDox^0}   & -r_1 & \dfrac{w_1}{\tk_L} \\
     0                   & 0    &   0  & - \dfrac{1}{\tDox^0}  & 0    & 0     \\
   - p_2                 & 0    & -q_2 & \dfrac{u_2}{\tDox^0}   & -r_2 & \dfrac{w_2}{\tk_L}  \\
     0                   & 0    &  0   &  0      & 0    & - \dfrac{1}{\tk_L}   \\
   - p_3                 & 0    & -q_3 &  0      & -r_3 & 0    \\
\end{pmatrix}
   \label{eq:matA}
\end{equation}
For example, the elements of the first two rows in $\matA$ are clear from Eq.\eqref{eq:teta1x}
written in terms of the $\vecv$ elements:
\begin{equation}
   \begin{split}
       &\pdr{\teta^1}{\tx} = - \dfrac{\tN_p^1}{\tsion^0} \\
       &\pdr{\tN_p^1}{\tx}  = - p_1 \teta^1 - q_1 \tc^1 + \dfrac{u_1}{\tDox^0}\pdr{\tN_{ox}^1}{\tx}
           - r_1 \tp_L^1 + \dfrac{w_1}{\tk_L}\pdr{\tN_L^1}{\tx}.
   \end{split}
   \label{eq:tNp}
\end{equation}

The matrix $\matA$ is constant on $[a, b]$, which allows us to
write a formal solution to Eq.\eqref{eq:eqnA}:
\begin{equation}
   \vecv(\tx) = \lexp{\matA (\tx - b)}\vecv_b, \quad a \leq \tx \leq b
   \label{eq:solform}
\end{equation}
Setting here $\tx = a$ we get the state vector $\vecv_a$ 
\begin{equation}
   \vecv_a = \matM_{ab}\vecv_b
   \label{eq:vecy_sol}
\end{equation}
where $\matM_{ab}$ is the required transform matrix 
\begin{equation}
   \matM_{ab} = \lexp{- \matA d}, \quad d = b - a > 0
   \label{eq:Mab}
\end{equation}
We postpone the discussion of numerical implementation of Eq.\eqref{eq:Mab}
to Section \ref{sec:num}.

\subsection{Impedance}

Consider now the calculation of the CCL impedance. Suppose for a moment that the elements of
the state vector $\vecv_1$ at $\tx=1$ are fully defined:
\begin{equation}
    \vecv_1 = \left(\teta_1^1, \tN_p^1, \tc_1^1, \tN_{ox,1}^1, \tp_{L,1}^1, \tN_{L,1}^1 \right)^T.
    \label{eq:v1}
\end{equation}
 The vector $\vecv_2$
at the interface between cells 1 and 2 (Figure \ref{fig:sketch}) is
$\vecv_2 = \matM_1 \vecv_1$, where $\matM_1$ is the transform
matrix $\vecv_1 \to \vecv_2$.
The vector $\vecv_3$ at the cells 2/3 interface
is $\vecv_3 = \matM_2 \vecv_2 = \matM_2\matM_1\vecv_1$. Continuing this process, we get
the state vector $\vecv_{N+1}$ at the membrane/CCL interface:
\begin{equation}
   \vecv_{N+1} = \matU \vecv_1,
   \label{eq:vN1}
\end{equation}
where $\matU$ is the transform matrix from $\tx=1$ to $\tx=0$:
\begin{equation}
   \matU = \matM_{N}\matM_{N-1}\ldots \matM_1
   \label{eq:matU}
\end{equation}

With $\vecv_{N+1}$ at hand, the perturbation of the cell current density at $\tx=0$ is given by
\begin{multline}
   \tj_{N+1}^1 = \left(- \tsion^0\pdr{\teta^1}{\tx} - \tsion^1\pdr{\teta^0}{\tx}\right)_{\tx=0} \\
     = \left(- \tsion^0\pdr{\teta^1}{\tx}  + \dfrac{\tj_0}{\tsion^0} \ksig\ks (\tp_L^1 - \tc^1)\right)_{\tx=0} \\
     = (\vecv_{N+1})_2 + \dfrac{\tj_0}{\tsion^0} \ksig\ks \bigl((\vecv_{N+1})_5 - (\vecv_{N+1})_3\bigr)
     \label{eq:tj1N}
\end{multline}
where the numeric subscripts denote vector $\vecv_{N+1}$ components and the dimensionless constants
$\ksig$, $\ks$ are given in Appendix~\ref{sec:app}.
The system impedance is
\begin{equation}
   \tZ = \dfrac{\teta^1_{N+1}}{\tj^1_{N+1}} = \dfrac{(\vecv_{N+1})_1}{\tj_{N+1}^1}
   \label{eq:tZ}
\end{equation}
With Eq.\eqref{eq:vN1}, the impedance, Eq.\eqref{eq:tZ},  is fully determined by the product $\matU\vecv_1$.

Standard BVP solvers use iterations to find solutions satisfying the boundary conditions
on either side of the interval  $[0, 1]$.
The main advantage of the matrix method is Eq.\eqref{eq:vN1}, which provides the linear
algebraic equations between $\vecv_1$ and $\vecv_{N+1}$ components. This allows us to calculate
the missing components of $\vecv_1$, the vector $\vecv_{N+1}$ and the impedance,
Eq.\eqref{eq:tZ}, without iterations.
The calculation of the $\vecv_1$ components is described in Appendix \ref{sec:bvp}.

\subsection{Numerical implementation}
\label{sec:num}

The key operation of the matrix method is
the calculation of the matrix $\matM$ for each numerical cell,
which includes the matrix exponent, Eq.\eqref{eq:Mab}.
This is done using the spectral decomposition of the matrix $\matA$:
\begin{equation}
   \matM = \lexp{- \matA d} = \left(\matr{E} \lexp{\matr{- \Lambda}d }\right)\matr{E}^{-1}
   \label{eq:spectral}
\end{equation}
where $\matr{E}$ contains six column eigenvectors of $\matA$,
and $\matr{\Lambda}$ is the diagonal matrix with the six $\matA$ eigenvalues.
The Python NumPy subroutine {\em np.linalg.eig(A)} for calculation of eigenvalues
and eigenvectors of the square matrix does the job. Furthermore,
the NumPy subroutines support batch matrix
exponential and product, for all the frequencies at once.
Neither the calculation of $\lexp{- \matA d}$, nor the  product of $N$ matrices
$\matM_n$, Eq.\eqref{eq:matU},
require a loop over the frequencies, which leads to a very efficient code.

However, there is a pitfall when using Eq.\eqref{eq:spectral}. At the frequency of around 500 Hz,
the resulting $\matU$ matrix, Eq.\eqref{eq:matU}, produces the discontinuity
in the impedance spectrum. The reason for that could be a degeneracy of at least one of the $\matA$
matrices at this frequency, meaning that the two eigenvalues of $\matA$ nearly coincide.
In that case, $\matA$ cannot be diagonalized and the algorithm of matrix exponentiation fails.
To resolve the problem, the matrix $\matA$ elements with
$r_1, w_1, r_2, w_2, p_3, q_3, r_3$ describing the liquid water transport
were multiplied by the filtering Heaviside function
\begin{equation}
   H(f-f_*) = \left\{ \begin{split}
                       & 1, \quad f < f_*  \\
                       & 0, \quad f \geq f_*
                      \end{split}
              \right.
    \label{eq:heavy}
\end{equation}
where $f_* = 200$ Hz is the threshold frequency.  At the frequencies above $f_*$,
the contribution of liquid water transients
to the cell impedance is vanishingly small, so this
filtering of $\matA$ elements does not distort the impedance.
However, the filtering  changes the spectral properties of $\matA$ stabilizing the calculation
of the matrix exponent.

\subsection{Cell segmentation and the procedure of impedance calculation}

To incorporate oxygen transport in the straight cathode channel,
the cell is separated into $N_{seg}$ segments.
 Below, $\tz_m$ indicates the coordinate of the $m$th segment along the channel.
Calculation of the segment impedance includes the following steps.
\begin{enumerate}

\item Solve the steady-state equation for the local current
    density $\tj^0(\tz_m)$ (Ref.\cite{Kulikovsky_19a})
    and  Eq.\eqref{eq:tch0z} for the oxygen concentration
    in the channel $\tc_h^0(\tz_m)$ as discussed in Appendix \ref{sec:ssp}.

\item Solve the static equations \eqref{eq:teta0x}--\eqref{eq:tcv0x} using the
    standard BVP solver.

\item For each segment:

    \begin{itemize}
       \item Calculate the matrix $\matA$ elements using
           Eqs. \eqref{eq:pqrs3a}--\eqref{eq:pqrs2b}.
           Multiply $r_1, w_1, r_2, w_2, p_3, q_3, r_3$
           by the Heaviside filter, Eq.\eqref{eq:heavy}.

       \item Using the loop over $N$ computational cells,
            $n=1, 2, \ldots, N$ calculate the matrices $\matM_n$, Eq.\eqref{eq:spectral},
            and accumulate their product to get $\matU$, Eq.\eqref{eq:matU}.

        \item Calculate the oxygen concentration perturbation in the channel
            $\tc_h^1(\tz_m)$, Eq.\eqref{eq:tch1z_sol}, and then
            the components of the vector $\vecv_1$, Eqs.\eqref{eq:tN11}, \eqref{eq:tNL1}.
            To a first approximation, the perturbation of the liquid water flux from the
            membrane $\tN_{L,mem}^1$ can be set to zero. Alternatively,
            any model for the membrane impedance can be used to calculate $\tN_{L,mem}^1$.

        \item Calculate the vector $\vecv_{N+1}$, Eq.\eqref{eq:vN1}.

        \item Calculate the perturbation of current, Eq.\eqref{eq:tj1N}, and
            the segment impedance $\tZ_{seg,m}$, Eq.\eqref{eq:tZ}.

    \end{itemize}

\item  Calculate the cell impedance
   \begin{multline}
      Z_{cell} = \left(\dfrac{1}{N_{seg}}\sum_{m=1}^{N_{seg}} \dfrac{1}{Z_{seg,m}}\right)^{-1} \\
         + R_{HFR} + \ri\omega L_{cab} S_{cell}
      \label{eq:Zcell}
   \end{multline}
   where $R_{HFR}$ is the high-frequency resistivity, $L_{cab}$ is the cable inductance, and
   $S_{cell}$ is the cell active area.
\end{enumerate}

\section{Results and discussion}

\begin{figure}
\begin{center}
    \includegraphics[scale=0.45]{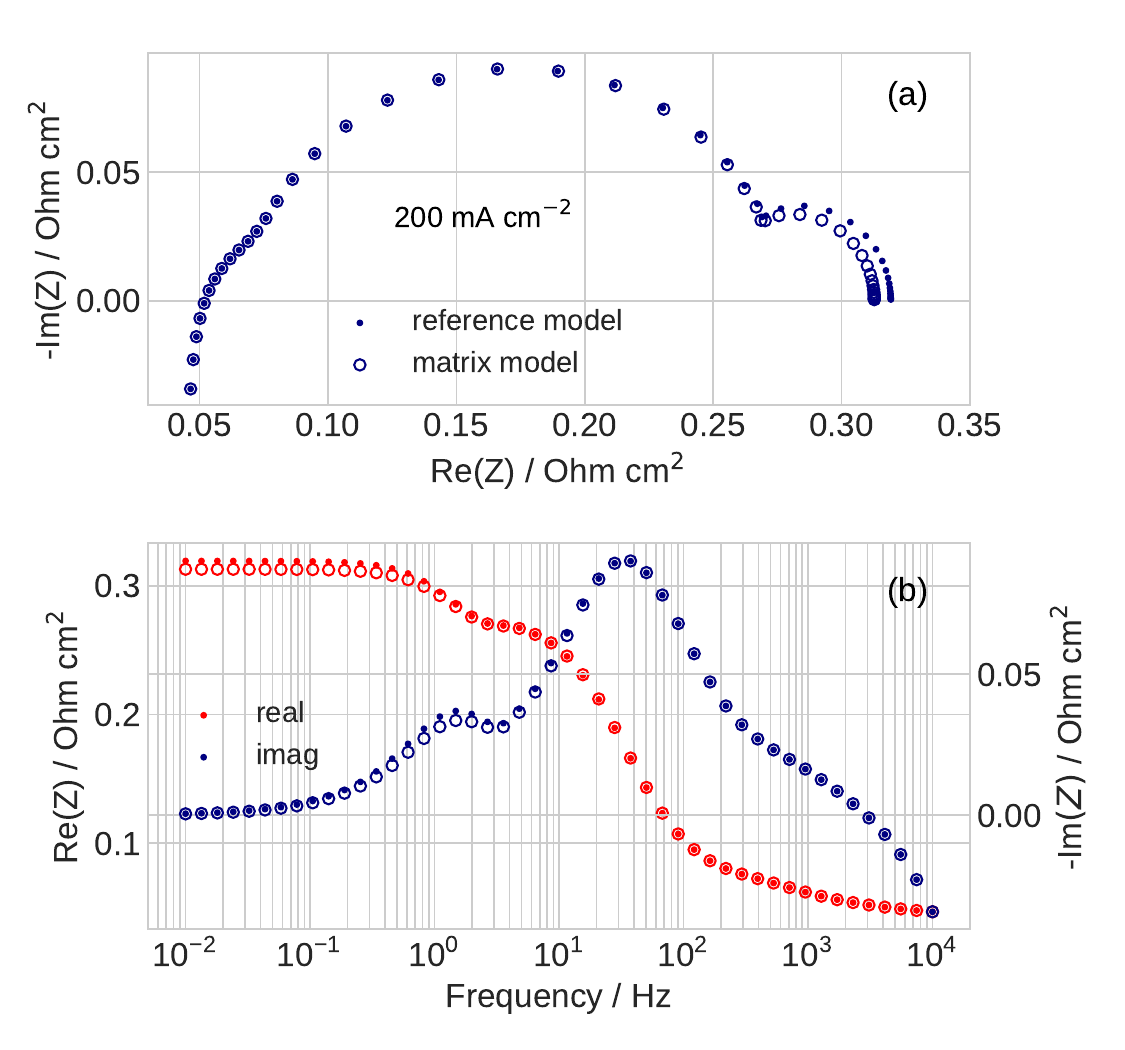}
\end{center}
\caption{(a) The Nyquist and (b) Bode plots of the PEMFC cathode impedance
    calculated using the reference model based of the Python BVP solver (points)
    and the matrix model (open circles) with the numerical mesh containing 10 points.
    The parameters used in the calculations are gathered in Table~\ref{tab:parms}.
    The matrix model returns the spectrum 150 times faster, than the reference model.
    }
\label{fig:s200}
\end{figure}

Figure~\ref{fig:s200} shows the Nyquist and Bode plots of the spectra calculated
using the matrix model and the reference model based on the general-purpose
Python BVP solver {\em solve\_bvp}.
The cell current density is 200 mA~cm$^{-2}$; the other operating and transport
parameters are indicated in Table~\ref{tab:parms}.

As can be seen, in the frequency range above 3 Hz,
the spectrum produced by the matrix model is nearly identical with the
reference one. A small gap between the spectra below 3 Hz is
due to the different algorithms for the approximate solution
of the plug-flow Eq.\eqref{eq:tch1z} (Appendix \ref{sec:bvp}).
In the reference model, Eq.\eqref{eq:tch1z} is solved using  $\tN_{ox,1}^1$
taken from the solution of the BVP for the perturbation amplitudes in every segment.
In the matrix model, Eq.\eqref{eq:tch1z_sol} is solved first,
and then $\tN_{ox,1}^1$ is obtained from Eq.\eqref{eq:tNox1}.

In this example, the matrix model with 10 numerical cells
returns the spectrum 200 times faster than the reference model. The poor
performance of the reference model is due to
(i) internal iterations in the finite element solver {\em solve\_bvp},
(ii) the loop over the frequencies, since {\em solve\_bvp} does not support batch operations,
and (iii) iterations to obtain the self-consistent solution of Eq.\eqref{eq:tch1z}.
The matrix model is essentially free from these loops. Instead, it includes
the batch calculations of the $\matM$ matrices using the spectral decomposition of the matrix $\matA$,
and the product of the $\matM$ matrices to give the $\matU$ matrix.
These procedures are much faster than the three nested loops in the reference model.
It is worth noting that the spectral decomposition works well for the complex-valued
matrices, while the standard BVP solver requires to convert the system of complex-valued
ODEs to equations for the real and imaginary parts.

\begin{table}
\small
\begin{center}
\begin{tabular}{|l|c|}
        \hline
        CCL oxygen diffusivity $\Dox$, cm$^2$s$^{-1}$ &  $2.1\cdot 10^{-4}$  \\
        CCL water vapor diffusivity $D_v$, cm$^2$s$^{-1}$ &  $5\cdot 10^{-3}$  \\
        ORR Tafel slope, V / exp             &  0.03     \\
        ORR exchange current density         &          \\
        $i_*$, A~cm$^{-3}$                   &  $10^{-4}$   \\
        Double layer capacitance $\Cdl$, F~cm$^{-3}$  & 20  \\
        CCL proton conductivity $\sion$, S~cm$^{-1}$  & $11\cdot 10^{-3}$ \\
        GDL oxygen diffusivity $D_b$, cm$^2$s$^{-1}$ &  $0.02$  \\
        \hline
        CCL thickness $\lcat$, cm            &  $10\cdot 10^{-4}$     \\
        GDL thickness $l_b$, cm              &   $235\cdot 10^{-4}$   \\
        Number of segments $N_{seg}$         &   8                    \\
        \hline
        Cathode absolute pressure, bar       &  1.5   \\
        Cathode flow RH                      &  0.5   \\
        Air flow stoichiometry $\lam$        &  4     \\
        Cell temperature, K                  &  273 + 80 \\
        Cell current density $j_0$, mA~cm$^{-2}$   & 200 \\
        \hline
        CCL hydraulic permeability\cite{Yufan_23} $K_L$, m$^2$     & $0.398\cdot 10^{-18}$  \\
        Water molar volume $V_m$, cm$^3$~ mol$^{-1}$ & 18          \\
        Water dynamic viscosity@$80^\circ$ $\mu_w$, Pa~s    &  $3.54\cdot 10^{-4}$  \\
        The elementary charge   $e_{0}$, C    & $1.6 \cdot 10^{- 19}$   \\
        Evaporation rate constant $\kappa$, Pa$^{- 1}$ m$^{- 2}$s$^{- 1}$    & $1.38\cdot 10^{17}$   \\
        Evaporation interfacial area factor $\xi_{ev}$, &  $80$  \\
        \hline
\end{tabular}
\end{center}
   \caption{Cell parameters used in the calculation
       of the spectra in Figure~\ref{fig:s200}.
    }
   \label{tab:parms}
\end{table}

\begin{figure}
\begin{center}
    \includegraphics[scale=0.45]{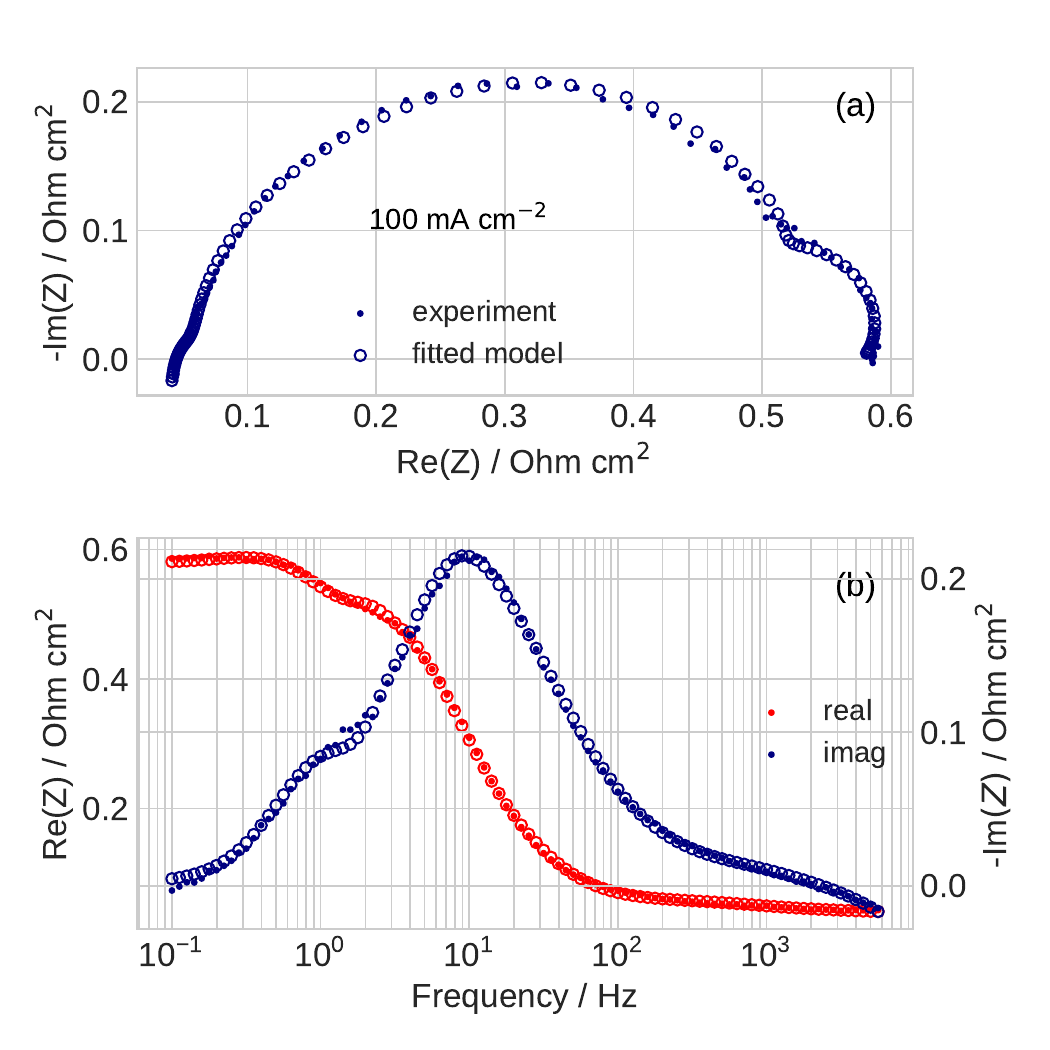}
\end{center}
\caption{(a) The experimental (solid points) and fitted matrix model (open circles)
     Nyquist spectra of a PEM fuel cell operated at the current density of 100 mA~cm$^{-2}$.
     The list of operating parameters is collected in Table~\ref{tab:fitted}.
     (b) The Bode plots of the real and imaginary parts of the spectra in (a).
    }
\label{fig:fitted}
\end{figure}

The matrix model fitted to the experimental spectrum of a PEM fuel cell
is shown in Figure~\ref{fig:fitted}. The cell current density is 100 mA~cm$^{-2}$;
the other operating parameters are shown in Table~\ref{tab:parms}.
The impedance spectrum was measured with 20 points per decade; the other experimental
details can be found in \cite{Kulikovsky_26b}.

Note the low-frequency ``hook'' in the Nyquist spectrum (Figure~\ref{fig:fitted}a)
caused by the dependence of CCL proton conductivity
on the slow relaxation of the liquid water saturation in the CCL.
The fitting parameters are collected in Table~\ref{tab:fitted}. Their numerical values
agree well with our previous measurements \cite{Kulikovsky_26b}, except the lower liquid saturation.
Here, however, for simplicity the perturbation of water flux from
the membrane was set to zero, while in Ref.\cite{Kulikovsky_26b} the transient
water flux from the membrane to the CCL was taken into account. Fitting using custom Python code took 60 seconds
on a notebook with an eight-core, 4.4 GHz AMD processor. For this problem, the matrix model
was 63 times faster than the reference model.

%


%
\begin{table}
\small
\begin{center}
\begin{tabular}{|l|c|}
        \hline
        ORR Tafel slope, mV / exp                      & $30.7 \pm 0.3$   \\
        Double layer capacitance $\Cdl$, F~cm$^{-3}$   & $24.9 \pm 0.3$   \\
        CCL proton conductivity $\sion$, mS~cm$^{-1}$  & $14 \pm 4$       \\
        CCL oxygen diffusivity $\Dox$, cm$^2$s$^{-1}$  & $(0.22 \pm 0.02)\cdot 10^{-4}$  \\
        Mean liquid saturation $s$                     & $0.31 \pm 0.01$   \\
        Capillary pressure $p_{c,0}$, bar              & $0.030 \pm 0.008$  \\
        WRT slope $k_{pc}$, bar                        & $0.141 \pm 0.006$ \\
        HFR, mOhm~cm$^2$                               & $35   \pm 1$   \\
        \hline
        Cell current density $j_0$, mA~cm$^{-2}$       & 100  \\
        Fitting code runtime on a laptop, s            & 50   \\
        \hline
\end{tabular}
\end{center}
\caption{Fitted parameters for the spectrum in Figure~\ref{fig:fitted} and the runtime
         of the fitting code.
}
   \label{tab:fitted}
\end{table}

\begin{figure}
\begin{center}
    \includegraphics[scale=0.45]{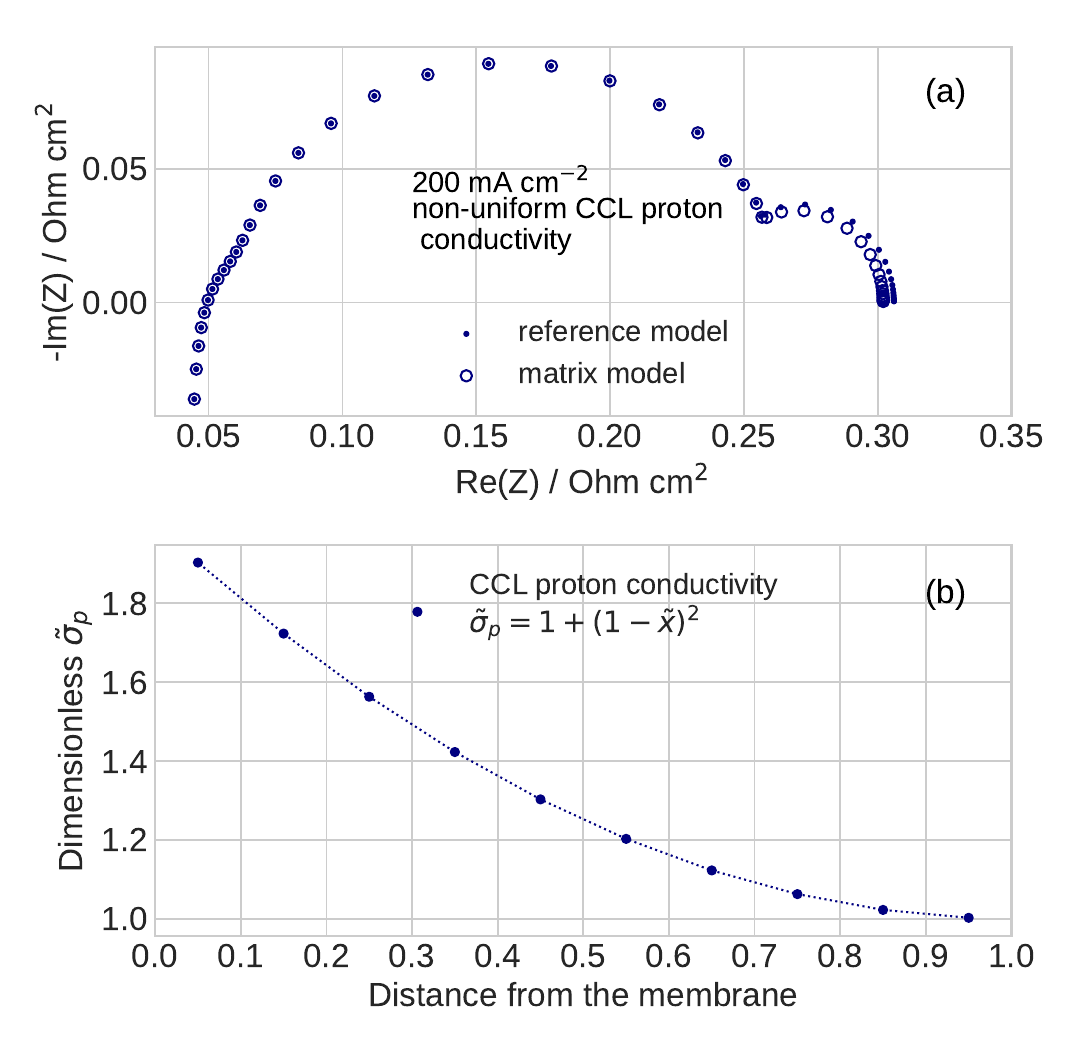}
\end{center}
\caption{(a) The Nyquist plots of the PEMFC cathode impedance
    calculated using the reference and matrix models.
    The parameters used in the calculations are the same as in Figure~\ref{fig:s200},
    except the CCL proton conductivity. (b) The normalized shape of the CCL proton conductivity
    at uniform CCL saturation.
    }
\label{fig:sgrad}
\end{figure}

The matrix model is suitable for describing impedance of functionally graded catalyst layers.
Figure~\ref{fig:sgrad} shows the spectra calculated for the same set of parameters
as the spectra in Figure~\ref{fig:s200}, except the proton conductivity, which at uniform
liquid saturation increases parabolically toward the membrane (Figure~\ref{fig:sgrad}b).
Comparison of Figures~\ref{fig:s200}a and \ref{fig:sgrad}a shows that the proton
conductivity gradient in Figure \ref{fig:sgrad}b reduces the cell polarization resistance.

The matrix model can be implemented for the system of linear ODEs of any order.
It requires some analytical work to construct the matrix $\matA$
for the system to be solved and to calculate the missing
elements of the initial state vector $\vecv_1$.
However, the game worth the candles, as this approach gives two orders
of magnitude faster code as compared to the general-purpose finite-element BVP solver.

Note that the shapes of the static
variables $\teta^0$, $\tc^0$, and $\tp_L$, which appear in the elements of the matrix
$\matA$ should be calculated using the general-purpose BVP solver. At present, application
of the matrix method for solution of nonlinear BVPs has not been studied.


\section{Conclusions}

A matrix model for the fast calculation of PEM fuel cell impedance was developed. In addition
to the proton and oxygen transport processes within the cell, the model takes into account
the transport of liquid water in the cathode catalyst layer (CCL) and water
evaporation. The liquid water transients affect the CCL oxygen diffusivity
and proton conductivity, leading to a low-frequency ``inductive'' loop in the Nyquist spectra.

The impedance model includes three linear second-order ODEs with variable coefficients,
which form a boundary-value problem on the interval $\tx \in [0,1]$. To solve this problem,
we introduce a numerical mesh on this interval. On each numerical cell we consider
the respective Cauchy
problem with constant coefficients. The Cauchy problem
is solved numerically using the spectral decomposition of the coefficient matrix.
The solution to the BVP on the interval $[0,1]$ is then assembled from
these solutions. Generally, this method allows  a linear BVP with variable
coefficients of an arbitrary order to be solved.

The model is two orders of magnitude faster than the reference model based
on the general-purpose Python BVP solver {\em solve\_bvp}.
The model is suitable for calculating the impedance of functionally graded catalyst layers.
Fitting the model to an experimental spectrum with 20 points per decade
takes 50 seconds on a laptop.

\appendix

\section{Derivation of Eqs.\eqref{eq:teta1x}--\eqref{eq:tpL1x}}
\label{sec:app}

Assuming that the static transport parameters $\tsion^0$ and $\tDox^0$ are constant
on a small numerical cell $[a,b]$,
linearization and Fourier-transform of equations \eqref{eq:etax}--\eqref{eq:pLx}
lead to the system of equations for the small perturbation amplitudes
$\teta^1$, $\tc^1$, and $\tp_L^1$ (Ref.\cite{Kulikovsky_24b}):
\begin{multline}
   \veps^2 \tsion^0\pddr{\teta^1}{\tx}
       = \left(\eeo\tc^0 + \ri\tom\tCdl\right)\teta^1 + \eeo\tc^1 \\
          - \veps^2\pdr{}{\tx}\left(\tsion^1\pdr{\teta^0}{\tx} \right)
    \label{eq:teta1xa}
\end{multline}
\begin{multline}
   \veps^2 \tDox^0\pddr{\tc^1}{\tx}
          = \eeo\tc^0\teta^1 + \left(\eeo + \ri\tom\mu^2\right)\tc^1  \\
             - \veps^2\pdr{}{\tx}\left(\tDox^1\pdr{\tc^0}{\tx} \right)
     \label{eq:tcx1a}
\end{multline}
\begin{equation}
   \tk_L\pddr{\tp_L^1}{\tx} = - 2\eeo\tc^0\teta^1 - 2\eeo\tc^1 + \ri\tom\gamma^2 s^1
   \label{eq:tpL1xa}
\end{equation}

Here, $\veps$, $\mu$, $\tk_L$ and $\gamma$ are the constant parameters:
\begin{multline}
    \veps = \sqrt{\dfrac{\sionast b_T}{i_* \lcat^2}}, \quad \mu = \sqrt{\dfrac{4 F\cref}{\Cdl b_T}}, \\
    \tk_L = \dfrac{4 F R T \cref K_L}{i_*\lcat^2\mu_w V_m}, \quad
   	\gamma = \sqrt{\dfrac{4 F}{V_m  \Cdl b_T}}.
     \label{eq:tkL}
\end{multline}

The perturbations $\tsion^1$, $\tDox^1$, $s^1$  are related to $\tp_L^1$ and $\tc^1$:
\begin{equation}
   \begin{split}
      & s^1     = \pdr{s^0}{p_c}\tp_c^1 = \ks \left(\tp_L^1 - \tc^1\right), \quad   \\
      &\tsion^1 = \ksig s^1 = \ksig\ks \left(\tp_L^1 - \tc^1\right) \\
      &\tDox^1  = \pdr{\tDox^0}{s}s^1 = \kD s^1  = \kD\ks \left(\tp_L^1 - \tc^1\right), \\
    \end{split}
   \label{eq:s1}
\end{equation}
where $\tp_L^1 - \tc^1 = \tp_c^1$ is the perturbation of the capillary pressure,
$\ksig$ is the dimensionless proportionality coefficient between the liquid saturation
and $\tsion$ : $\tsion = \ksig s$,
and the following notations are introduced
\begin{multline}
    \ks \equiv \pdr{s^0}{p_c} \\
    = \dfrac{(1 - s_*)}{2}\left(1 - \tanh^2\left(\dfrac{p_c^0 - p_{c,0}}{k_{pc}}\right) \right)
         \dfrac{R T \cref}{k_{pc}}
         \label{eq:tks}
\end{multline}
\begin{equation}
    \kD \equiv \pdr{\tDox^0}{s} = - 2 (1 - s^0) \tDoxd.
    \label{eq:tkD}
\end{equation}
Here
\begin{equation}
   p_c^0 = p_L^0 - p_g^0 = p_L^0 - (p_{ox}^0 + p_{N_2}^0 + p_v^0)
   \label{eq:pc0}
\end{equation}
and $p_{N_2}^0$ is the static nitrogen pressure.
Note that the dimensionless oxygen concentration and pressure perturbations are equal:
$\tc^1 = \tp_{ox}^1$ and $\tc_v^1 \equiv 0$.

Substituting Eqs.\eqref{eq:s1} into the system \eqref{eq:teta1xa}--\eqref{eq:tpL1xa} we get
\begin{multline}
   \veps^2 \tsion^0\pddr{\teta^1}{\tx}
       = \left(\eeo\tc^0 + \ri\tom\tCdl\right)\teta^1 + \eeo\tc^1 \\
          - \veps^2\ksig\ks\pdr{}{\tx}\left(\left(\tp_L^1 - \tc^1\right)\pdr{\teta^0}{\tx} \right)
    \label{eq:teta1xb}
\end{multline}
\begin{multline}
   \veps^2 \tDox^0\pddr{\tc^1}{\tx}
          = \eeo\tc^0\teta^1 + \left(\eeo + \ri\tom\mu^2\right)\tc^1  \\
             - \veps^2\kD\ks\pdr{}{\tx}\left(\left(\tp_L^1 - \tc^1\right)\pdr{\tc^0}{\tx} \right)
     \label{eq:tc1xb}
\end{multline}
\begin{equation}
   \tk_L\pddr{\tp_L^1}{\tx} = - 2\eeo\tc^0\teta^1 - 2\eeo\tc^1
          + \ri\tom\gamma^2 \ks\left(\tp_L^1 - \tc^1\right)
   \label{eq:tpl1xb}
\end{equation}

Expanding the derivatives in the last term in Eqs.\eqref{eq:teta1xb}, \eqref{eq:tc1xb}, we find
\begin{multline}
   \veps^2\tsion^0\pddr{\teta^1}{\tx}
       = \left(\eeo\tc^0 + \ri\tom\right)\teta^1 + \eeo\tc^1 \\
          - \veps^2\ksig\ks\left(\pdr{\tp_L^1}{\tx}\pdr{\teta^0}{\tx} + \tp_L^1\pddr{\teta^0}{\tx}\right) \\
          + \veps^2\ksig\ks\left(\pdr{\tc^1}{\tx}\pdr{\teta^0}{\tx} + \tc^1\pddr{\teta^0}{\tx}\right)
    \label{eq:teta1xc}
\end{multline}
\begin{multline}
    \veps^2\tDox^0\pddr{\tc^1}{\tx}
          = \eeo\tc^0\teta^1 + \left(\eeo + \ri\tom\mu^2\right)\tc^1  \\
           - \veps^2\kD\ks\left(\pdr{\tp_L^1}{\tx}\pdr{\tc^0}{\tx} + \tp_L^1\pddr{\tc^0}{\tx}\right) \\
           + \veps^2\kD\ks\left(\pdr{\tc^1}{\tx}\pdr{\tc^0}{\tx} + \tc^1\pddr{\tc^0}{\tx}\right).
     \label{eq:tcx1c}
\end{multline}
Collecting like terms, we get the system \eqref{eq:teta1x}--\eqref{eq:tpL1x} with
the coefficients
\begin{multline}
    p_1 = \dfrac{\tc^0\eeo + \ri\tom\tCdl}{\veps^2}, \quad
    q_1 = \dfrac{\eeo}{\veps^2} + \ksig\ks \pddr{\teta^0}{\tx}, \\
    r_1 = - \ksig\ks \pddr{\teta^0}{\tx}, \quad
    u_1 = \ksig\ks \pdr{\teta^0}{\tx},    \quad
    w_1 = - u_1
    \label{eq:pqrs1}
\end{multline}
\begin{multline}
    p_2 = \dfrac{\tc^0\eeo}{\veps^2}, \quad
    q_2 = \dfrac{\eeo + \ri\tom\mu^2}{\veps^2} + \kD\ks \pddr{\tc^0}{\tx}, \\
    r_2 = - \kD\ks \pddr{\tc^0}{\tx}, \quad
    u_2 = \kD\ks \pdr{\tc^0}{\tx}, \quad
    w_2 =  - u_2
    \label{eq:pqrs2}
\end{multline}
\begin{equation}
    p_3 = - 2\tc^0\eeo, \quad
    q_3 = - 2\eeo - \ri\tom\gamma^2\ks  \quad
    r_3 = \ri\tom\gamma^2\ks
    \label{eq:pqrs3a}
\end{equation}

Finally, using the static equations $\pddra{\teta^0}{\tx} = \tc^0\eeo/(\veps^2\tsion^0)$,
$\pddra{\tc^0}{\tx} = \tc^0\eeo/(\veps^2\tDox^0)$,
we find the complete set of parameters for the matrix $\matA$ elements
\begin{multline}
    p_1 = \dfrac{\tc^0\eeo + \ri\tom\tCdl}{\veps^2}, \quad
    q_1 = \dfrac{\eeo}{\veps^2} + \ksig\ks \dfrac{\tc^0\eeo}{\veps^2\tsion^0}, \\
    r_1 = - \ksig\ks \dfrac{\tc^0\eeo}{\veps^2\tsion^0}, \quad
    u_1 = \ksig\ks \pdr{\teta^0}{\tx}, \quad
    w_1 = - u_1
    \label{eq:pqrs1b}
\end{multline}
\begin{multline}
    p_2 = \dfrac{\tc^0\eeo}{\veps^2}, \quad
    q_2 = \dfrac{\eeo + \ri\tom\mu^2}{\veps^2} + \kD\ks \dfrac{\tc^0\eeo}{\veps^2\tDox^0}, \\
    r_2 = - \kD\ks \dfrac{\tc^0\eeo}{\veps^2\tDox^0}, \quad
    u_2 = \kD\ks \pdr{\tc^0}{\tx}, \quad
    w_2 =  - u_2
    \label{eq:pqrs2b}
\end{multline}
with $p_3, q_3$, and $r_3$ given by Eq.\eqref{eq:pqrs3a}.
Note that all the parameters in Eqs.\eqref{eq:pqrs1b}, \eqref{eq:pqrs2b}, \eqref{eq:pqrs3a}
are calculated
in the middle of a numerical cell. It is recommended to convert each frequency-independent
parameter to an array of constant elements for using in batch subroutines of Python.

\section{Steady-state problem}
\label{sec:ssp}

The static shapes $\teta^0$, $\tc^0$, $\tp_L^0$, and $\tc_v^0$ obey the system
of static equations
\begin{multline}
   \veps^2\pdr{}{\tx}\left(\tsion^0\pdr{\teta^0}{\tx}\right) = \tc^0\expo, \\
    - \left(\tsion^0\pdr{\teta^0}{\tx}\right)_{\tx=0} = \tj^0,
       \quad \left.\pdr{\teta^0}{\tx}\right|_{\tx=1} = 0
      \label{eq:teta0x}
\end{multline}
\begin{equation}
   \tDox^0\pdr{\tc^0}{\tx} = \tj^0 + \tsion^0\pdr{\teta^0}{\tx}, \quad \tc^0(1) = \tc_1^0
      \label{eq:tc0x}
\end{equation}
\begin{multline}
   - \tk_L \pddr{\tp_L^0}{\tx} = 2\tc^0\expo - \tk_{ev} (\tc_v^s - \tc_v^0), \\
   - \left(\tk_L\pdr{\tp_L^0}{\tx}\right)_{\tx=0} = 4\veps^2 \alpha_m \tj^0, \quad \tp_L^0(1) = \tp_{cell}
   \label{eq:tpL0x}
\end{multline}
\begin{multline}
   -\veps^2\tD_v\pddr{\tc_v^0}{\tx} = \tk_{ev} (\tc_v^s - \tc_v^0), \\
   \left.\pdr{\tc_v^0}{\tx}\right|_{\tx=0} = 0, \quad \tc_v^0(1) = \tc_v^s R_H,
   \label{eq:tcv0x}
\end{multline}
where the oxygen concentration at the CCL/GDL interface $\tc_1^0$ is
\begin{equation}
   \tc_1^0 = \tc_h^0(\tz) - \dfrac{\tj^0}{\tjlim}, \quad \jlim = \dfrac{4 F \cref D_b}{l_b},
   \label{eq:tc1}
\end{equation}
$\tj^0$ is the local current density in the segment,
$\alpha_m$ is the effective water transfer coefficient through the membrane, $0 \leq R_H \leq 1$ is the
relative humidity of the cathode flow, and
\begin{equation}
   \tk_{ev} = \dfrac{4F\cref e_0\kappa \xi_{ev} RT  }{\lcat i_*}.
   \label{eq:kec}
\end{equation}
The shape of the local current density $\tj^0(\tz)$ along the cathode channel can be
calculated using the model \cite{Kulikovsky_19a}. With $\tj^0(\tz)$ at hand,
the oxygen concentration along the channel is determined
from the dimensionless static plug-flow Eq.\eqref{eq:chz}:
\begin{equation}
   \lam\tJ \pdr{\tc_h^0}{\tz} = - \tj^0(\tz), \quad \tc_h^0(0) = 1,
   \label{eq:tch0z}
\end{equation}
where $\tJ$ is the mean current density in the cell, $\lam$ is the air flow stoichiometry.
The local current density in Eqs.\eqref{eq:teta0x}--\eqref{eq:tpL0x} is, therefore,
$\tj^0 = \tj^0(\tz)$.

Eq.\eqref{eq:tcv0x} is separated from the system with the solution
\begin{multline}
   \tc_v^0 = \tc_v^s \left(1 - \left(1 - R_H\right)
               \dfrac{\cosh\left(\phi\tx\right)}
                     {\cosh\phi} \right), \\
                     \quad \phi \equiv \sqrt{\tk_{ev}/(\veps^2\tD_v)},
    \label{eq:tcv0sol}
\end{multline}
which gives us the explicit rate of evaporation in Eq.\eqref{eq:tpL0x}:
\begin{equation}
   \tR_{ev} = \tk_{ev} (\tc_v^s - \tc_v^0)
            = \tk_{ev} \tc_v^s \left(1 - R_H\right)
               \dfrac{\cosh\left(\phi\tx\right)}{\cosh\phi}.
\end{equation}

\section{Boundary conditions and the vector $\vecv_1$}
\label{sec:bvp}

Eq.\eqref{eq:vN1} relating the state vectors $\vecv_1$ at $\tx=1$ and $\vecv_{N+1}$ at $\tx=0$
allows us to solve the boundary-value problem with any set of boundary conditions on either
side of the CCL.
For definiteness, consider the system  Eqs.\eqref{eq:teta1x2}--\eqref{eq:tpL1x2}
with the variable static transport parameters.


Two components of $\vecv_1$, Eq.\eqref{eq:v1},
are zero: the proton current perturbation $\tN_{p,1}^1 = 0$, and
the liquid pressure perturbation $\tp_{L,1}^1 = 0$. The second condition
means that we ignore the transport of liquid water in the GDL.
The perturbation $\teta_1^1$ is fixed, it can be any small real number.
The perturbation $\tc_1^1$ follows from the solution of the oxygen transport
equation in the GDL and channel (see below). The fluxes $\tN_{ox,1}^1$
and  $\tN_L^1$ at $\tx=1$ have to be determined using the boundary conditions at $\tx=0$.

We have two boundary conditions on the membrane side ($\tx=0$):
the perturbation of the oxygen flux is zero,
and the perturbation of the liquid water flux is continuous.
Thus, from Eq.\eqref{eq:vN1} we get
\begin{equation}
    \begin{split}
        &(\vecv_{N+1})_4 = (\matU \vecv_1)_4 = 0 \\
        &(\vecv_{N+1})_6 = (\matU \vecv_1)_6 = \tN_{L,mem}^1
    \end{split}
    \label{eq:flux1_mem}
\end{equation}
where
$\tN_{L,mem}^1$ is the perturbation amplitude of the water flux incoming from the membrane.
Eqs.\eqref{eq:flux1_mem} are two linear algebraic equations for  $\tN_{ox,1}^1$ and $\tN_L^1$:
\begin{equation}
    \begin{split}
        & U_{4,1}\teta_1^1 + U_{4,3}\tc_1^1 + U_{4,4}\tN_{ox,1}^1 + U_{4,6}\tN_{L,1}^1 = 0 \\
        & U_{6,1}\teta_1^1 + U_{6,3}\tc_1^1 + U_{6,4}\tN_{ox,1}^1 + U_{6,6}\tN_{L,1}^1 = \tN_{L,mem}^1
    \end{split}
    \label{eq:flux1_eqns}
\end{equation}
Solution of the oxygen transport equation in the GDL leads
to the relation of the oxygen concentration
perturbation $\tc_1^1$ at the CCL/GDL interface
with that perturbation in the channel $\tc_h^1$ (Ref.\cite{Kulikovsky_17i})
\begin{equation}
   \tc_1^1 = \alpha \tN_{ox,1}^1 + \beta \tc_h^1,
   \label{eq:tc11}
\end{equation}
where $\alpha$,  $\beta$ are the coefficients that take into account oxygen transport
in the GDL:
\begin{multline}
   \alpha = \tanh\left(\mu\tl_b\theta\right)/\left(\mu\tD_b\theta\right), \quad
                  \beta = 1/\cosh\left(\mu\tl_b\theta\right), \\
            \theta \equiv \sqrt{\ri\tom/\left(\veps^2\tD_b\right)}
   \label{eq:albeta}
\end{multline}
Here,
$\tD_b$ and $\tl_b$ are the dimensionless GDL oxygen diffusivity and thickness.

Substituting Eq.\eqref{eq:tc11} into Eq.\eqref{eq:flux1_eqns}, and solving this system, we get
\begin{multline}
   \tN_{ox,1}^1 =  - \dfrac{\beta\tc_h^1}{d_N}(U_{43} U_{66} - U_{46} U_{63})  \\
              - \dfrac{\teta_1^1}{d_N}(U_{41} U_{66} - U_{46} U_{61})
              - \dfrac{\tN_{L,mem}^1 U_{46}}{d_N}
   \label{eq:tN11}
\end{multline}
\begin{multline}
   \tN_L^1 = \dfrac{\beta\tc_h^1}{d_N}(U_{43} U_{64} - U_{44} U_{63}) \\
             + \dfrac{\teta_1^1}{d_N}(\alpha (U_{41} U_{63} - U_{43} U_{61})
                 + U_{41} U_{64} - U_{44} U_{61}) \\
             + \dfrac{\tN_{L,mem}^1}{d_N}(\alpha U_{43} + U_{44}),
    \label{eq:tNL1}
\end{multline}
where
\begin{equation}
   d_N = \bigl(U_{44} U_{66} - U_{46} U_{64}\bigr) +  \alpha\bigl(U_{43}U_{66} - U_{46} U_{63}\bigr).
   \label{eq:dN}
\end{equation}

To complete the vector $\vecv_1$ calculation,
we need a solution to the equation for the oxygen concentration
perturbation $\tc_h^1$ in the channel \cite{Kulikovsky_17i}:
\begin{equation}
   \lam\tJ \pdr{\tc_h^1}{\tz} =  - \zeta\tc_h^1 + \beta\tN_{ox,1}^1, \quad \tc_h^1(0) = 0,
   \label{eq:tch1z}
\end{equation}
where $\zeta$ is the constant parameter
\begin{equation}
   \zeta = \ri\tom\xi^2 + \mu\sqrt{\ri\tom\tD_b/\veps^2}
                          \tanh\left(\mu \tl_b \sqrt{\ri\tom / \left(\veps^2\tD_b\right)} \right)
   \label{eq:zeta}
\end{equation}
and
\begin{equation}
   \xi = \sqrt{\dfrac{4 F h \cref\lcat i_*}{\Cdlast \sionast b^2}}.
   \label{eq:xi}
\end{equation}
Eq.\eqref{eq:tch1z} follows from the Fourier transform
of the plug-flow equation, Eq.\eqref{eq:chz}, with the
diffusive oxygen flux on the right side
resulting from the GDL transport equation  \cite{Kulikovsky_15g,Kulikovsky_19d}.

It is convenient to write Eq.\eqref{eq:tN11} in the form
\begin{equation}
    \tN_{ox,1}^1 = P_h \tc_h^1 + Q_h + R_h,
    \label{eq:tNox1}
\end{equation}
where 
\begin{equation}
   \begin{split}
      & P_h = - \dfrac{\beta}{d_N}(U_{43} U_{66} - U_{46} U_{63})      \\
      & Q_h = - \dfrac{\teta_1^1}{d_N}(U_{41} U_{66} - U_{46} U_{61})  \\
      & R_h = - \dfrac{\tN_{L,mem}^1 U_{46}}{d_N}.
   \end{split}
   \label{eq:PQR}
\end{equation}

The parameters $P_h$, $Q_h$, and $R_h$ depend on the segment position
and the solution to Eq.\eqref{eq:tch1z}
is given by the following recurrence relation with $m$ enumerating the segments:
\begin{multline}
   \tc_{h,m}^1 = \left(\tc_{h,m-1}^1 + B_m\right)
      \lexp{\dfrac{(\beta P_{h,m} - \zeta)\td_z}{\lam\tJ}} - B_m, \\
      B_m \equiv \dfrac{\beta(Q_{h,m} + R_{h,m})}{\beta P_{h,m} - \zeta}
      \label{eq:tch1z_sol}
\end{multline}
Assuming that the oxygen concentration perturbation at the channel inlet is zero,
the initial condition for Eq.\eqref{eq:tch1z_sol} in the first segment at
the distance $\td_z/2$ from the channel inlet is  ($m=1$)
\begin{multline}
   \tc_{h,1}^1 = \left(0 + B_1\right)
      \lexp{\dfrac{(\beta P_{h,1} - \zeta)\td_z}{2\lam\tJ}} - B_1, \\
      B_1 \equiv \dfrac{\beta(Q_{h,1} + R_{h,1})}{\beta P_{h,1} - \zeta}
      \label{eq:tch1z_sol0}
\end{multline}
Here $\td_z$ is the dimensionless segment length along the $\tz$-axis.
The local $\tN_{ox,1}^1(\tz_m)$ in the $m$th
segment is obtained using Eq.\eqref{eq:tch1z_sol} in Eq.\eqref{eq:tNox1}.
With Eqs.\eqref{eq:tN11}, \eqref{eq:tNL1}, \eqref{eq:tch1z_sol}, and \eqref{eq:tNox1}
the state vector $\vecv_1$ is fully determined.

Note that $\tc_h^1$, $\tN_{ox}^1$, and $\tc_1^1$ are expressed through
the matrix $\matU$ elements. In impedance models
based on standard BVP solvers, the $\matU$  matrix is absent
and the above parameters are calculated by iterating the solutions to the oxygen transport
problems along the channel and in the through-plane direction. This incurs additional
overhead in terms of CPU time.

\newpage


\newpage

\section*{Nomenclature}

\small

\begin{tabular}{ll}
    $\tilde{}$   &  Marks dimensionless variables                             \\
    $\matA$      &  ODEs system matrix, Eq.\eqref{eq:matA}                     \\
    $a,b$        &  Boundaries of the numerical cell                          \\
    $B_m$        &  Dimensionless parameter in Eq.\eqref{eq:tch1z_sol}        \\
    $b_T$        &  ORR Tafel slope, V                                        \\
    $c$          &  Oxygen concentration in the CCL, mol~cm$^{-3}$           \\
    $c_b$        &  Oxygen concentration in the GDL, mol~cm$^{-3}$           \\
    $c_h$        &  Oxygen concentration in the channel, mol~cm$^{-3}$           \\
    $\cref$      &  Reference oxygen concentration, mol cm$^{-3}$        \\
    $c_v$        &  Water vapor molar concentration, mol~cm$^{-3}$           \\
    $\Cdl$       &  Double layer capacitance,$F$ cm$^{-3}$       \\
    $D_b$        &  Oxygen diffusivity in the GDL, cm$^2$~s$^{-1}$   \\
    $\Dox$       &  Oxygen diffusivity of the CCL, cm$^2$~s$^{-1}$   \\
    $D_{ox,dry}$   &  Oxygen diffusivity of the dry CCL, cm$^2$~s$^{-1}$   \\
    $D_v$        &  Water vapor diffusivity of the CCL, cm$^2$~s$^{-1}$   \\
    $d$          &  Length of the computational cell $d= b - a$, cm                   \\
    $e_{0}$      &  Elementary charge, $C$                            \\
    $F$          &  Faraday constant, C~mol$^{-1}$                            \\
    $h$          &  Channel height, cm    \\
    $\ri$        &  Imaginary unit                                            \\
    $i_*$        &  Volumetric exchange current density, A~cm$^{-3}$                     \\
    $J$          &  Mean cell current density, A~cm$^{-2}$                     \\
    $j$          &  Local proton current density in the CCL, A~cm$^{-2}$       \\
    $j_*$        &  Characteristic current density, A~cm$^{-2}$       \\
    $K_L$        &  CCL liquid water permeability, m$^{2}$       \\
    $k_{pc}$     &  Slope of the water retention curve, Pa       \\
    $k_{\sigma}$ &  Proportionality factor between saturation \\
                 &  and proton conductivity, Eq.\eqref{eq:Doxs}, S~cm$^{-1}$       \\
    $k_L$        &  Liquid transport coefficient, mol m$^{-1}$ Pa $^{-1}$ s$^{-1}$  \\
    $\tk_D$      &  Dimensionless coefficient, Eq.\eqref{eq:tkD}                 \\
    $k_{ev}$     &  Evaporation  coefficient, mol m$^{-3}$ Pa $^{-1}$ s$^{-1}$       \\
    $\tk_s$       &  Dimensionless coefficient, Eq.\eqref{eq:tks}                 \\
    $\lcat$      &  CCL thickness, m                                         \\
    $l_b$        &  GDL thickness, m                                         \\
    $\matM$      &  Transfer matrix from $\tx=b$ to $\tx=a$, Eq.\eqref{eq:Mab} \\
    $N$          &  Number of computational cells along the $\tx$-axis       \\
    $N_{seg}$    &  Number of segments along the cathode channel              \\
    $\tN_L^1$    &  Dimensionless  perturbation of \\
                 &  liquid water flux, Eq.\eqref{eq:bcxb}   \\
    $\tN_{L,mem}^1$    &  Dimensionless perturbation of \\
                 &  liquid water flux from membrane, Eq.\eqref{eq:tpL1x2}    \\
    $\tN_{ox}^1$ &  Dimensionless perturbation of \\
                 &   oxygen flux, Eq.\eqref{eq:bcxb}             \\
    $\tN_p^1$    &  Dimensionless perturbation of \\
                 &  proton current density, Eq.\eqref{eq:tNp}   \\
    $p_v^s$      &  Saturated vapor pressure at 80°C \\
                 &  at flat liquid-gas interface, bar                                  \\
    $p_c$        &  Capillary pressure, Pa               \\
    $p_L$        &  Liquid pressure in the CCL, Pa                \\
    $p_{N_2}$    &  Nitrogen pressure in the CCL, Pa                \\
    $p_v$        &  Vapor pressure in the CCL, Pa               \\
\end{tabular}

\begin{tabular}{ll}
    $p_{ox}$     &  Oxygen pressure in the CCL, Pa               \\
    $p_{cell}$   &  Cell gaseous pressure, Pa                              \\
    $P_h,Q_h,R_h$ &  Dimensionless parameters, Eq.\eqref{eq:PQR}   \\
    $R$          &  Gas constant, $\text{J}/(\text{K}\ \text{mol})$                \\
    $R_H$        &  Relative humidity,     $0 \leq R_H \leq 1$            \\
    $R_{HFR}$    &  High-frequency (ohmic) resistance, Ohm~cm$^2$            \\
    $R_{ev}$     &  Rate of water evaporation A~cm$^{-3}$    \\
    $R_{ORR}$    &  ORR rate, A~cm$^{-3}$                              \\
    $s$          &  CCL liquid saturation, $0 \leq s \leq 1$                  \\
    $s_*$        &  Minimal liquid saturation, $0 \leq s_* \leq 1$                  \\
    $T$          &  Cell temperature, K                                      \\
    $t$          &  Time, s                                                  \\
    $t_*$        &  Characteristic time, s                                   \\
    $\matU$      &  Transfer matrix from $\tx=1$ to $\tx=0$, Eq.\eqref{eq:matU} \\
    $V_m$        &  Molar volume of liquid water, m$^{3}$mol$^{- 1}$       \\
    $\vecv$      &  State vector, Eq.\eqref{eq:vecv}                        \\
    $x$          &  Coordinate through the cell,m       \\
    $Z$          &  Electrochemical impedance, Ohm cm$^{2}$              \\
    $z$          &  Coordinate along the cathode channel, cm       \\ [1em]
\end{tabular}

{\bf Subscripts:}

\begin{tabular}{ll}
    $0$      & membrane/CCL interface \\
    $1$      & CCL/GDL interface  \\
    $b$      & GDL  \\
    $h$      & Channel  \\
    $L$      & Liquid water   \\
    $m$      & Membrane; molar volume \\
    $*$      & Characteristic or reference value \\ [1em]
\end{tabular}

{\bf Superscripts:\\}

\begin{tabular}{ll}
    $0$      & Steady--state value \\
    $1$      & Small--amplitude perturbation \\
    $*$      & Characteristic, or reference value \\ [1em]
\end{tabular}

{\bf Greek:\\}

\begin{tabular}{ll}
    $\alpha$, $\beta$   &  Dimensionless parameters, Eq.\eqref{eq:albeta}\\
    $\gamma$, $\veps$   &  Dimensionless parameters, Eq.\eqref{eq:tkL}\\
    $\eta$              &  ORR overpotential, positive by convention, V \\
    $\zeta$             &  Dimensionless parameter, Eq.\eqref{eq:zeta}   \\
    $\mu$               &  Dimensionless parameter, Eq.\eqref{eq:tkL}   \\
    $\mu_w$             &  Dynamic viscosity of liquid water, Pa~s \\
    $\kappa$            &  Evaporation rate constant, atm$^{- 1}$ m$^{- 2}$s$^{- 1}$  \\
    $\xi$               &  Dimensionless parameter, Eq.\eqref{eq:xi}        \\
    $\xi_{\text{ec}}$   &  Evaporation interfacial area factor  \\
    $\rho$              &  Air density, kg~m$^{-3}$                     \\
    $\sion$             &  CCL proton conductivity, S~m$^{-1}$        \\
    $\sionast$          &  Reference CCL proton conductivity, S~m$^{-1}$        \\
    $\phi$              &  Dimensionless parameter, Eq.\eqref{eq:albeta} \\
    $\omega$            &  Angular frequency of the AC signal, s$^{-1}$
\end{tabular}

\newpage

\end{document}